\documentclass[journal]{vgtc}

\title{A Multiplexing Design Space: Theory, Method, and Application}

\author{%
  \authororcid{Yiwen Xing}{0000-0003-1521-6616},
  \authororcid{Afrah Farea}{0000-0003-4412-5377},
  \authororcid{Saiful Khan}{0000-0002-6796-5670}, and 
  \authororcid{Min Chen}{0000-0001-5320-5729}
}

\authorfooter{
  \item
  	Yiwen Xing is with University of Oxford.
  	E-mail: yiwen.xing@eng.ox.ac.uk.
  \item
  	Afrah Farea is with Istanbul Technical University.
  	E-mail: afrah.farea@outlook.com.
  \item Saiful Khan is with Science and Technology Facilities Council (STFC).
  	E-mail: saiful.khan@stfc.ac.uk.
  \item
  	Min Chen is with University of Oxford.
  	E-mail: min.chen@oerc.ox.ac.uk.  
}

\abstract{%
Many visualization designs feature phenomena referred to as ``visual multiplexing'', where multiple pieces of information associated with the same data point are conveyed simultaneously. Although visualization designers are able to bring such phenomena, often unconsciously, into their designs, the design space of visual multiplexing is huge, and it is uncommon to explore visual multiplexing systematically as design patterns.
In this paper, we propose a design method for exploring a smaller design space constrained by an application. As an illustrative case study, we focus on machine learning (ML) workflows for developing ML models that approximate partial differential equations (PDEs). In these workflows, ML researchers need to analyze the inter-relationships among multiple 2D scalar fields frequently.
Since superimposing one heatmap on top of another is not an effective design, we formulate three design steps to explore the design space of visual multiplexing in the context of multiple 2D scalar fields. Our design method also includes a pre-design step for domain grounding and theoretical analysis, and involves domain experts in both co-design and evaluation activities. The design process enables us to identify relatively optimal default multiplexing designs as well as the need for small variations that domain experts can control through a user interface.  
}


\graphicspath{{figs/}{figures/}{pictures/}{images/}{./}}

\usepackage{tabu}                     
\usepackage{booktabs}                 
\usepackage{array}
\usepackage{longtable}
\usepackage{mathptmx}                  
\usepackage{amsmath}
\usepackage{amssymb}
\usepackage[dvipsnames]{xcolor}
\usepackage{cite}
\usepackage{comment}
\begin{document}

\firstsection{Introduction}
\label{sec:Intro}

\maketitle

\begin{figure*}[h]
  \centering
  \includegraphics[width=\linewidth]{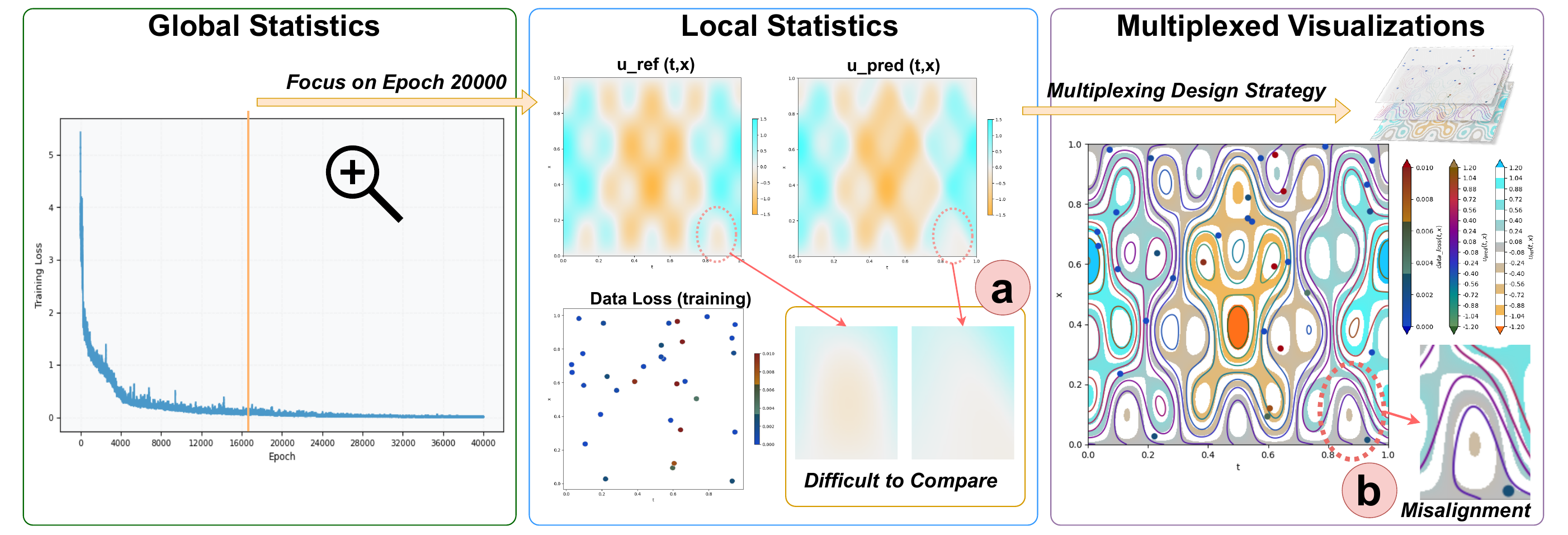}\par
  \caption{%
  	Global statistics, local statistics recorded in three fields, and a multiplexed visualization of the three fields. (a) and (b) show the same pair of data (\textit{u\_ref} and \textit{u\_pred}) using juxtaposition and superposition, respectively, where the latter employs contour lines over a zebra heatmap. This comparison illustrates how multiplexed representations can better support visual comparison between fields.%
  }
  \label{fig:teaser}
\end{figure*}

\emph{Visual multiplexing} is a collection of common phenomena in data visualization, where multiple pieces of information associated with the same data point are conveyed simultaneously. Chen et al.~\cite{Chen:2014:CGF} borrowed the term ``multiplex'' from the field of tele- and data-communication, and categorized these phenomena into 12 types. While the first two types correspond directly to space- and time-division multiplexing in communication, 
\textcolor{black}{the other ten types rely on human perceptual and cognitive capabilities to decode multiple signals related to data at a single location in a plot. For example, \emph{partial occlusion} allows one visual object to be perceived despite being partially covered by another; \emph{translucent occlusion} enables viewers to interpret overlapping objects through transparency; and \emph{integrated visual channels} encode multiple attributes within the same visual object (e.g., using color and shape simultaneously).}
With visual representations featuring one or more phenomena of the latter ten types, humans can perceive, most of the time correctly, multiple pieces of information about the same data point, even when some information is spatially-shifted, visually-modified, geometrically-distorted, partly-occluded, totally-absent, or some other forms of impairment and deterioration,
\textcolor{black}{as studied in Gestalt psychology~\cite{koffka1935principles}}.
Chen et al. offered an information-theoretical explanation based on the information-theoretical framework for visualization \cite{Chen:2010:TVCG}.

\textcolor{black}{Although multiplexing is widely used in visualization practice, it is typically applied in an ad-hoc manner, without a systematic exploration of the underlying design space. Given the 12 types of multiplexing phenomena, the corresponding context-free design space becomes combinatorially large and difficult to navigate.}
However, with appropriate context, such as an application, data, users, tasks, and plot types, a context-specific design space can be explored systematically. For example, the application studied in this work aims to enhance visualization support in a \emph{machine learning} (ML) workflow for developing models that approximate mathematically-defined \emph{partial differential equations} (PDEs). The main data type handled by such an ML model is of 1D space plus time (1D+T), while other data to be visualized include various performance metrics captured during model training and testing. The users are ML researchers specialized in developing such models. Their primary task is to improve ML techniques used in developing models for different PDEs in a multi-cycle workflow that spans months or years. Their main visualization tasks are to observe the various data available in the ML workflow to inform their decisions on improving ML models and techniques. Because of the 1D+T main data type, they work with different types of 2D scalar fields, $F_1(x, t), F_2(x, t), \ldots$,  as illustrated in the center of Figure \ref{fig:teaser}. Due to difficulties in superimposing one heatmap on the top of another, they relied extensively on aggregating data in 2D scalar fields $F_1(x, t), F_2(x, t), \ldots$ into 1D global statistical measures $G_1(v), G_2(v), \ldots$ ($v$ is $x$ or $t$), allowing multiple $G_j(v)$ to be displayed using line plots as exemplified on the left of Figure \ref{fig:teaser}.

ML researchers often found that those line plots did not provide sufficient information for them to make decisions, e.g., what to do next after a training-and-testing run. It is highly desirable for users to observe the relationships among different 2D scalar fields, in addition to global statistical measures. Hence, this application provides a compelling case study for examining the methodology of exploring a multiplexing design space. 
\textcolor{black}{Rather than presenting a conventional design study~\cite{Sedlmair:2012:TVCG}, this work adopts a theory-guided perspective, where the application serves as an example to instantiate and examine a general method.}

As the overall contribution of this work,
\textcolor{black}{we propose a theory-guided method for systematically constructing a multiplexing design space under a specific application context---the PDE machine learning process. The method features two novel aspects (\textit{a} and \textit{b}), together with a supporting collaborative process (\textit{c}), as listed below:}
\begin{itemize}
    \item[a.] We introduce a pre-design step for analyzing the requirements based on the information-theoretical cost-benefit analysis \cite{Chen:2016:TVCG} (Section \ref{sec:Theory}),  \textcolor{black}{strengthening the traditional user-centered approach.}
    \item[b.] We formulate a three-step approach for exploring a multiplexing design space \textcolor{black}{related to an application} systematically (Section \ref{sec:Method}).%
    \item[c.] We engage with domain experts closely to consider, evaluate, and select different design options, \textcolor{black}{evidencing that the theory-guided approach complements the traditional approach.} (Section \ref{sec:CoDesign}).
\end{itemize}

\section{Related Work}

\noindent
\textbf{Machine Learning for Partial Differential Equations (PDEs).}
PDEs are traditionally solved using numerical methods, such as the Finite Element Method (FEM) or the Finite Difference Method (FDM). These approaches discretize a domain into mesh points and approximate the solution by solving the governing equations at those points. Lagaris, Likas, and Fotiadis~\cite{Lagaris:1998:TNN} first demonstrated the potential of neural networks for solving PDEs, leading to a research agenda followed by many computational scientists and machine learning (ML) researchers, e.g.,  \cite{Wang:2016:sandia,Wang:2017:PRF,Raissi:2019:JCP,Farea:2025:HiPC}. Because PDEs in practical applications are often nonlinear, high-dimensional, and lacking closed-form analytical solutions, FEM, FDM, and other numerical methods for solving PDEs are computationally demanding. The research on ML models to solve PDEs has shown the potential of ML models to reduce such costs significantly.
In particular, Physics-informed neural networks (PINNs) directly incorporate physical laws into the learning process by leveraging automatic differentiation to compute the required derivatives of the network’s output.
Such methods have shown advantageous potential for addressing the aforementioned challenges as they are mesh-free, scalable, and easily parallelizable, while offering continuous (closed-form) approximations that can be evaluated at arbitrary points in the domain.

Visualization plots are widely used in ML workflows, including those for developing ML models that approximate numerical PDE solvers. In particular, heatmaps are commonly used for visualizing errors and residuals. For example, a heatmap of absolute errors supports comparative analysis against traditional FEM and FDM solvers. A heatmap of domain-spanning residuals shows where the candidate solution diverges from the governing physical laws, providing a visual means for scientific validation. Localized residual information~\cite{Lu:2021:SIAM} and its distribution~\cite{Wu:2023:CMAME} can inform residual-driven adaptive sampling strategies, analogous to adaptive mesh refinement. Visualization has also been used to reveal internal network features, including activation maps and learned weight matrices (e.g.,~\cite{tensorboard}).

Beyond assessing training convergence and solution error, it is important for ML researchers to observe a range of performance metrics and training signals for rigorous assessment of ML processes as well as solutions. For example, Farea and Celebi \cite{Farea:2025:CPC} computed and visualized a broad range of performance metrics and training signals, including neural tangent kernel (NTK) analysis and Hessian eigenvalue decomposition.

\noindent
\textbf{Information Theory and Multiplexing.}
Chen and J\"{a}nicke proposed an information theoretic framework, providing quantitative measures and means of mathematical reasoning for explaining phenomena in visualization processes \cite{Chen:2010:TVCG}. They noticed a major difference between visualization and most other applications of information theory. In the latter (e.g., data communication), both encoding and decoding are machine-centric processes. In visualization, visual encoding is typically a machine-centric process, while visual decoding is a human-centric process. Chen et al. studied phenomena of visual multiplexing \cite{Chen:2014:CGF}, where multiple pieces of information associated with the same data point can be depicted simultaneously. With computer-generated visualization plots, visual multiplexing is realized by computers while de-multiplexing is achieved by humans. Chen et al. grouped these phenomena into 12 categories, identified several human factors that underpin humans' capability of de-multiplexing, e.g., Gestalt principles, memory, and knowledge, and used the information-theoretic framework \cite{Chen:2010:TVCG} to explain the mathematical validity of seemingly extra information conveyed beyond the bandwidth.
While designers may intuitively produce visual designs featuring visual multiplexing without the knowledge of information theory or the concept of visual multiplexing, one research question is: can we consciously make use of the concept of visual multiplexing by exploring its design space systematically? This is the focus of this paper.

Building on the information-theoretic framework, Chen and Golan discovered a mathematical formula that can explain, in abstraction, the trade-offs in many visualization and visual analytics techniques \cite{Chen:2019:CGF}. In Section \ref{sec:Theory}, we will use this formula to explain multiplexing, enriching the original validity explanation in \cite{Chen:2014:CGF}.
Chen and Ebert proposed a methodology for improving visual analytics workflow by analyzing the trade-offs among different machine- and human-centric processes \cite{Chen:2019:CGF}. It has been used to address shortcomings in visual designs as well as visual analytics systems (e.g., \cite{Ye:2023:TVCG,Jin:2024:TVCG}). In Section \ref{sec:Theory}, we applied this methodology to analyze the problems and potential solutions in our case study.

\noindent
\textbf{Visualization Design Space.}
A design space provides a structured way to enumerate and reason about the options available when constructing a visual design. By articulating the dimensions along which design decisions vary, a design space helps designers compare alternatives, understand trade-offs, and select appropriate encodings for a given analytical context~\cite{Elliott:2021:TVCG}. It also supports clearer communication and reproducibility by making implicit design choices explicit.

In visualization research, design spaces have been developed for various topics. Examples include design space for visualization tasks~\cite{Schulz:2013:TVCG}; multiscale visualizations~\cite{Solen:2025:VIS}; accessible visualizations~\cite{Kim:2021:CGF}, origin-destination data visualizations~\cite{Tennekes:2021:CGF}; visualization for games~\cite{Bowman:2012:TVCG}, and the visualization transformation from 2D data to 3D data~\cite{Lee:2022:CHI}. These design space explorations often focus on interaction techniques, visual encodings, or representational structures, enumerating the alternatives available within a particular design problem.

Work most closely related to ours is the design space of composite visualization by Javed and Elmqvist~\cite{Javed:2012:PVS}, which characterizes how multiple pieces of information can be combined within or across views using four high-level strategies: \textit{juxtaposition}, \textit{superimposition}, \textit{overloading}, and \textit{nesting}.
\textcolor{black}{Building on this framework, Deng et al. further surveyed design patterns of composite visualizations~\cite{Deng2023TVCG}, while Gleicher et al. examined the trade-offs between juxtaposition and superposition in the context of visual comparison~\cite{gleicher2018tvcg}.}
While these works provide an overarching understanding of visual composition, they operate at a relatively abstract level and do not examine the fine-grained design decisions required when multiple data attributes must be conveyed within a shared spatial domain.
\textcolor{black}{Layered representation is another relevant topic that has been extensively studied in cartography and geo-visualization, where maps are commonly constructed using base layers and overlays to integrate multiple sources of information (e.g.,~\cite{maceachren2004maps,Slocum2022}). These works highlight the importance of visual hierarchy and perceptual separability when combining multiple layers. These works resulted in map design principles focusing on what may work or not work. Our work provides a different methodology for exploring a combinatorial design space systematically, with layered encodings as an example. The systematical approach also enables a design process to explore the design space beyond layered encodings, for example, glyph encodings.
Multivariate encoding techniques such as glyph-based visualization~\cite{Borgo:2013:CGF} is a type of multiplexing, providing another means of depicting multiple attributes about a single data point. Within the unified framework of multiplexing, such non-layered approaches can integrate with a layered approach.}

Our work addresses the aforementioned gap by focusing on visual multiplexing --- the coordinated use of multiple visual channels within a single view to encode several attributes that share a common field. By identifying and organizing the multiplexing options relevant to these domain-coupled fields, we propose a more detailed design space that supports reasoning about the cost-benefit trade-offs of layering information within a unified view.

\begin{figure*}[t]
    \centering
    \includegraphics[width=\linewidth]{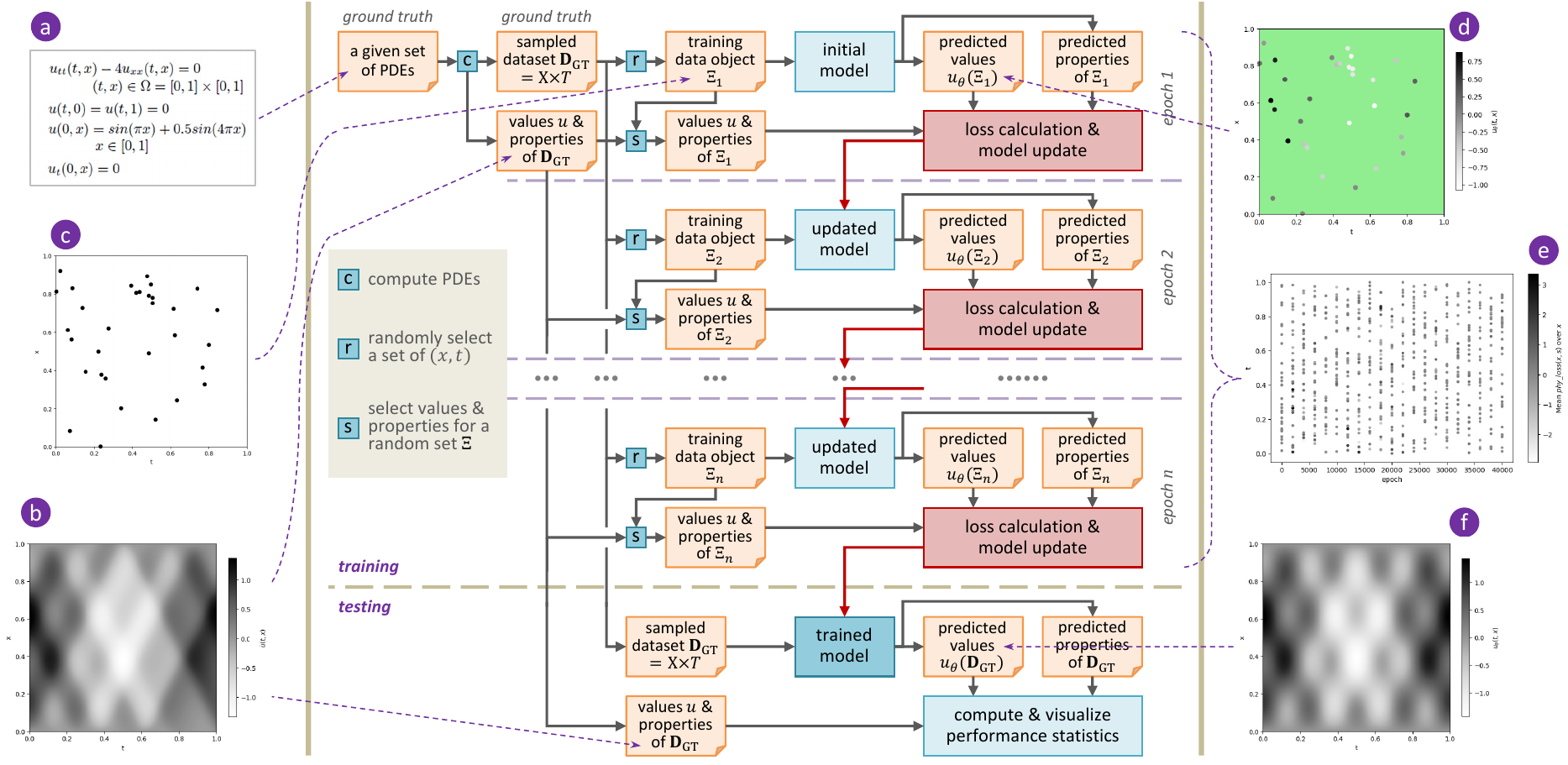}
    \caption{A machine learning (ML) workflow for training and testing an ML model that approximates a numerical PDEs solver.}
    \label{fig:MLworkflow}
\end{figure*}

\section{An Illustrative Application Case Study}
\label{sec:Context}

We consider a general form of PDEs:
\begin{align}
\mathcal{D}[u(\mathbf{x}); \alpha] &= f(\mathbf{x}), \quad \mathbf{x} \in \Omega, \\
\mathcal{B}_k[u(\mathbf{x})] &= g_k(\mathbf{x}), \quad \mathbf{x} \in \Gamma_k \subset \partial\Omega, \; k = 1,2, \ldots, n_b.
\end{align}
where $\mathcal{D}$ and $\mathcal{B}_k$ denote arbitrary differential and boundary (or initial) operators, respectively. The term $f(\mathbf{x})$ refers to the source term or forcing function. The functions $g_k(\mathbf{x})$ represent a set of boundary or initial condition values. $\alpha$ denotes a set of coefficients or parameters, and $u(\mathbf{x})$ is the solution of the differential equation given the input $\mathbf{x}$ in the solution domain $\Omega$ with boundary $\partial\Omega$.

As illustrated in Figure \ref{fig:MLworkflow}, given a specific set of PDEs \textcircled{a}, one can train a physics-informed neural network (PINN) to estimate the solution of the PDEs. A numerical solution $u(x, t)$ \textcircled{b} is first obtained, which is used as the ground truth throughout the ML workflow for developing this PINN. Figure \ref{fig:MLworkflow} depicts one training-testing run. The training process begins with a set of data points $\Xi_1$ randomly selected from a domain of $u(x, t)$ \textcircled{c}. The initial model attempts to predict the solution for the data points in $\Xi_1$, resulting in $u_\theta(x, t)$ at $\Xi_1$ \textcircled{d}. The loss of the first epoch is computed from $u(x, t)$ and $u_\theta(x, t)$ at $\Xi_1$, allowing the training process to update the initial model. The typical PINN loss function is defined as:
{\small
\begin{align*}
\mathcal{L}(\theta) &= \arg\min_{\theta} \sum_{k=1}^{n} \lambda_k \mathcal{L}_k(\theta) \\
&= \arg\min_{\theta} \left( \lambda_1 \mathcal{L}_1(\mathcal{D}[u_\theta(\mathbf{x}); \boldsymbol{\alpha}] - f(\mathbf{x})) + \sum_{k=2}^{n_b} \lambda_k \mathcal{L}_k(\mathcal{B}[u_\theta(\mathbf{x})] - g_k(\mathbf{x})) \right)
\end{align*}
}%
where $n_b$ denotes the total number of loss terms, $\mathcal{L}_k$ are the individual loss functions for each constraint with weighting coefficients $\lambda_k$, and $\theta$ represents the trainable parameters.

This process is repeated for many epochs as illustrated downwards in Figure \ref{fig:MLworkflow}. Different training signals of the model at different epochs can be measured and visualized. Figure \ref{fig:MLworkflow} \textcircled{e} shows a summary plot for one type of training signals phys\_loss.

Once the training process finishes, the trained model is tested, resulting in a predicted solution $u_\theta(x, t)$ for the whole domain $\Omega$ \textcircled{f}, which is evaluated against the numerical solution $u(x, t)$ using different performance metrics. The results of these performance metrics are typically visualized using heatmaps.

Images \textcircled{b}$\sim$\textcircled{f} in Figure \ref{fig:MLworkflow} represent the visualization of a few scalar fields in a training-testing run. There are many types of field data computed in the ML workflow.
In Appendix \ref{apx:DataSpace}, we list a collection of commonly-used fields identified by domain experts.
Furthermore, during the model development, ML researchers typically invoke the workflow numerous times in order to improve the model by experimenting with the model structure and different hyperparameters. Although these data fields can be visualized as heatmaps or their simplified variants (e.g., scatter plots for binary scalar fields), comparing and relating visual patterns inside different heatmaps is not easy when juxtaposing two or more heatmaps. Hence, domain experts rely more heavily on global statistics, as shown on the left of Figure \ref{fig:teaser}. At the beginning of this work, they expressed dissatisfaction about these plots for global statistics, hoping for better visualization to help them identify ways to improve a model more effectively.

\section{Information-Theoretical Analysis}
\label{sec:Theory}
\textcolor{black}{In this section, we introduce a theory-guided pre-design step that derives visualization requirements based on an information-theoretic cost-benefit analysis. Rather than relying solely on domain expert input, it enables designers to identify requirements that domain experts did not articulate due to insufficient knowledge of visualization. We will evidence the benefit of the theory-guided approach in Section \ref{sec:CoDesign}. 
}

\textbf{Visual Multiplexing.}
In communication, the concept of multiplexing was introduced around the middle of the 19th century \cite{Tucker:1971:REE}. Since visualization is a form of information transformation, there is some similarity between multiplexing in communication and visual multiplexing. There are four main types of multiplexing in communication, namely (a) space-, (b) time-, (c) frequency-, and (d) code-division multiplexing. Among the 12 types of visual multiplexing \cite{Chen:2014:CGF}, the first two types, ``A. partition a space'' and ``B. partition a time period'' are conceptually the same as (a) and (b) respectively. A few types of visual multiplexing feature combined signals on the same canvas, resembling (c). All 12 types of visual multiplexing feature different conventions or methods of visual encoding, resembling (d). However, there are two major differences. Firstly, in communication, a pair of multiplexer and de-multiplexer is developed together, while in visualization, visual designers do not have much control over de-multiplexers, i.e., how human viewers would interpret the multiplexed visual signals. Secondly, in communication, data compression is usually a separate process. For example, in the OSI 7-layer model for data communication, data compression is at layer 6, while multiplexing is at layer 1. In contrast, for most types of visual multiplexing, lossy data compression is an integral part, playing an essential role.

Together with the information-theoretic framework \cite{Chen:2010:TVCG}, the conceptualization of visual multiplexing influenced the formulation of the information-theoretic cost-benefit ratio \cite{Chen:2016:TVCG}, i.e.,
\[
    \frac{\textit{Benefit}}{\textit{Cost}} = \frac{\textit{Alphabet Compression }\text{(AC)} - \textit{Potential Distortion }\text{(PD)}}{\textit{Cost}}
\]
where AC measures the amount of entropy reduction (or information loss) achieved by a process, while PD balances the positive nature of AC by measuring the errors typically due to information loss. The denominator Cost provides a further balancing factor in addition to the trade-off between AC and PD. It includes the cost of the forward transformation from input to output (e.g., generating and viewing a multiplexed plot) and the inverse transformation for reconstructing the original input (e.g., interpreting a multiplexed plot, including de-multiplexing). Hence, a good use of visual multiplexing reflects a relatively optimal trade-off among AC, PD, and Cost. One can easily imagine that if we place one heatmap with 50\% opacity over another heatmap with 100\% opacity, it is a form of Type E multiplexing (``translucent occlusion''), but not very effective, as viewing each heatmap would be error-prone (high PD) and demand high cognitive load (high cost). In Section \ref{sec:Method}, we will examine a number of more effective multiplexing designs than combining two heatmaps naively.
For example, if a heatmap is approximated by a contour plot, which is then superimposed over a heatmap (Type G multiplexing---``depict a continuous ﬁeld''), this multiplexed plot would be of lower PD and lower Cost, while offering higher AC. Hence, the ratio of Benefit over Cost is higher.  

\noindent\textbf{ML Workflows for Modeling PDEs.}
The domain experts' dissatisfaction with global statistics in their workflows is not uncommon among ML researchers and developers, even though examining varied local statistics across multiple scalar fields occurs less often in ML workflows that lack spatiotemporal models. We adopted the information-theoretical guided methodology for improving visual analytics systems \cite{Chen:2019:CGF}, which has been used to improve ML workflows \cite{Ye:2023:TVCG,Jin:2024:TVCG,Saner:2025:AS}. Here we briefly describe the symptoms, causes, remedies, and side-effects identified in our analysis, and use $\mapsto$ to indicate the information-theoretic terms used in \cite{Chen:2019:CGF}. 
\begin{enumerate}
    \item \textbf{Symptom A:} Domain experts found that global statistics are not informative as to how to improve the current ML model being developed. [$\mapsto$ \textbf{High PD} in identifying issues or \textbf{High Cost} in hypothesizing possible solutions.]
    \item \textbf{Cause A:} Global statistics abstract information in an ML workflow quickly. [$\mapsto$ \textbf{High AC}.]
    \item \textbf{Remedy A:} Visualizing more local statistical patterns at the level of scalar fields. [$\mapsto$ \textbf{Reduce AC}.] One may consider a simple analogy: when observing the average performance of a whole school is not effective, one may observe the average performance of individual classes.
    \item \textbf{Side Effect A:} As local statistical patterns are in 2D scalar fields, visually analyzing many heatmaps is not easy, e.g., when relating different spatial patterns in juxtaposed heatmaps. [$\mapsto$ \textbf{High Cost} in terms of cognitive load \cite{Borgo:2010:TVCG}.]
    \item \textbf{Symptom B:} Make \textbf{Side Effect A} as \textbf{Symptom B}.
    \item \textbf{Cause B:} Displaying multiple heatmaps may not provide enough abstraction for domain experts to analyze multiple heatmaps with a reasonable cognitive load. [$\mapsto$ \textbf{Low AC}.]
    \textcolor{black}{A video is provided in the supplementary materials to demonstrate that juxtaposing and animating multiple heatmaps does not work well.}
    \item \textbf{Remedy B:} Using visual multiplexing, which introduces more abstraction [$\mapsto$ \textbf{Increase AC}] while reduces cognitive load for comparative spatial reasoning [$\mapsto$ \textbf{Reduce Cost}.]
    \item \textbf{Side Effects B:} (i) The design of visual multiplexing plots is not trivial, and (ii) domain experts will need to work with visual representations unknown to them before. To address (i), we explore the design space of multiplexing for an application systematically in order to identify relatively optimal designs (see Section \ref{sec:Method}). To address (i) and (ii), we engage domain experts in a co-design and evaluation process (see Section \ref{sec:CoDesign}).  
\end{enumerate}

\begin{figure*}[t]
    \centering
    \includegraphics[width=145mm]{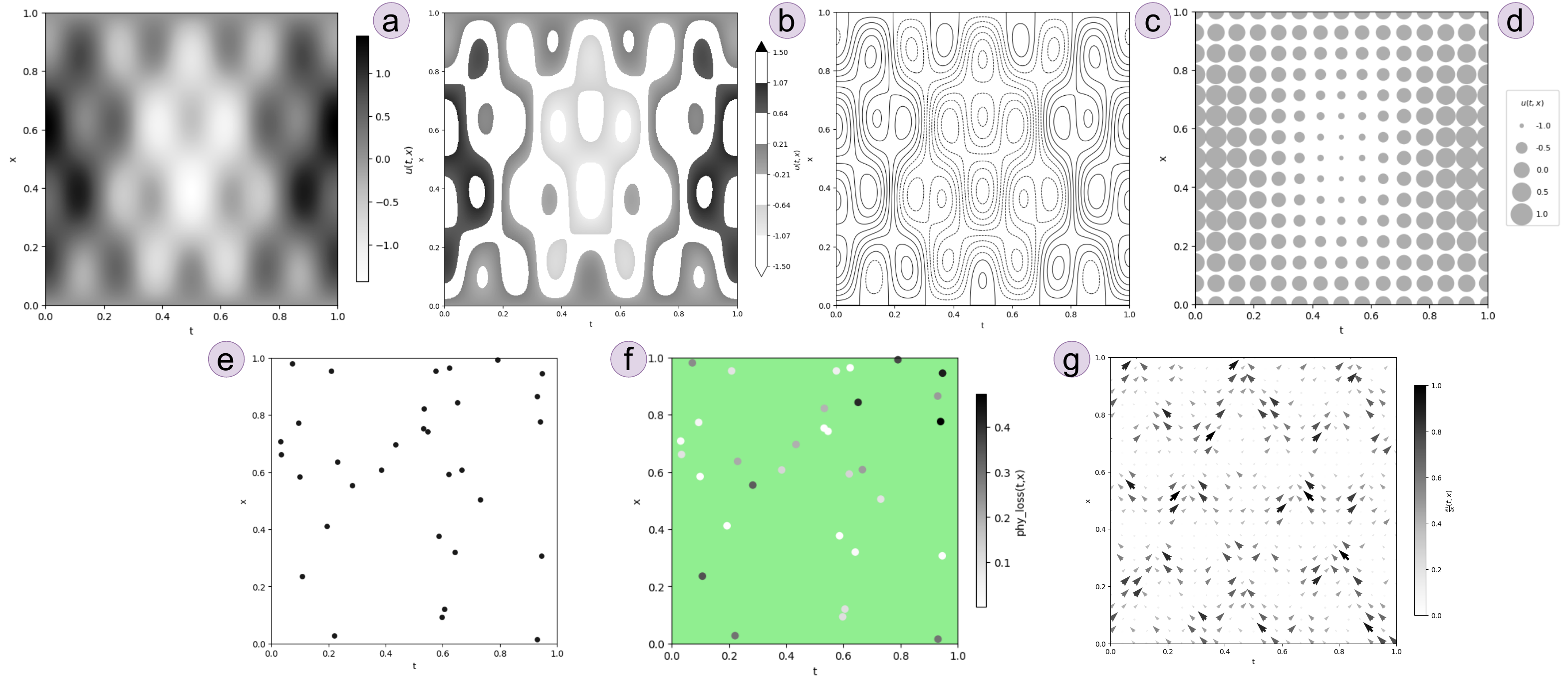}
    \caption{Examples of primitive design options: (a) \textbf{heatmap}; (b) \textbf{zebra map}; (c) \textbf{contour}; (d) \textbf{size glyph}; (e) \textbf{monochromic dots}; (f) \textbf{color glyph}; and (g) \textbf{pre-defined glyph}. Primitives (d)$\sim$(f) all belong to the \textbf{univariate glyph} family and can be applied either on a grid or at randomly sampled locations. Here, (d) illustrates a grid-based form, whereas (e) and (f) show randomly sampled instances.}
    \label{fig:Primitives}
\end{figure*}

\section{Constructing Multiplexing Design Space}
\label{sec:Method}
\textcolor{black}{In this section, we present a three-step method for constructing a multiplexing design space under a specific application context. While the method is application-independent, we use the case study described in Section \ref{sec:Context} to illustrate how the proposed method works in practice.}

\subsection{Exploring Context-Free Primitive Design Patterns}
\label{sec:Explore}
A \emph{primitive design pattern} is a \textbf{group} of design options that feature similar informational and visual properties. Figure \ref{fig:Primitives} shows several typical primitive design patterns for visualizing 2D field data. For example, the heatmap pattern encompasses all heatmaps with different colormaps, and they share some common characteristics, e.g., visual information is of a high density; typically, all data values in a discrete field are conveyed; each pixel conveys both its 2D location and the data value at the location; and usually it represents a scalar field (1D data).             

\begin{figure}[h]
    \centering
    \includegraphics[width=\linewidth]{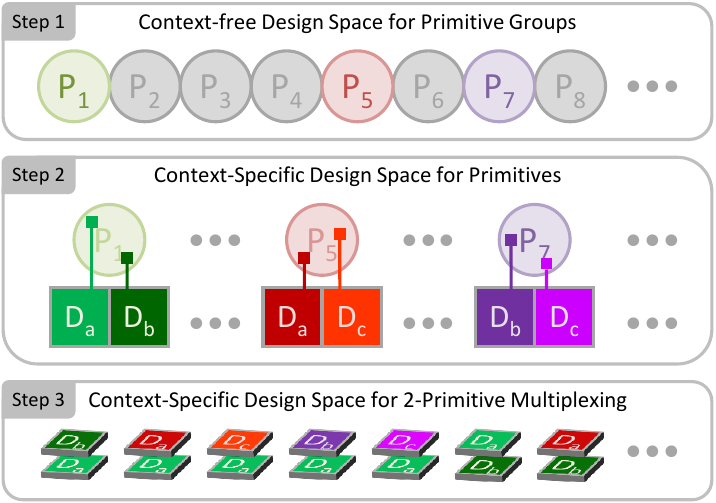}
    \caption{Three steps for constructing a multiplexing design space.}
    \vspace{-4mm}
    \label{fig:ThreeSteps}
\end{figure}

We can use the following four informational and visual properties to characterize a primitive design pattern:  

\begin{itemize}
    \item \emph{Information Density} (ID) --- It characterizes the ratio of the number of data points whose information is being depicted and an area. Coarsely, we can consider three levels, that is,
    \textbf{ID}$_1$: high density, e.g., (a) in Figure \ref{fig:Primitives};
    \textbf{ID}$_2$: medium density, e.g., (b) and (c); and \textbf{ID}$_3$: low density, e.g., (e) and (f).
    \item \emph{Sampling Strategy} (SS) --- It characterizes the methods for selecting data points to be displayed, including
    \textbf{SS}$_1$: all data points, e.g., (a) in Figure \ref{fig:Primitives};
    \textbf{SS}$_2$: structured sampling, e.g., (d);
    \textbf{SS}$_3$: data-guided, e.g., (b) and (c); and
    \textbf{SS}$_4$ random, e.g., (e) and (f).
    \item \emph{Information Type} (IT) --- There are three main types of information in the context of this paper, that is,
    \textbf{IT}$_1$: the location of each data point depicted;
    \textbf{IT}$_2$: the data value (or values) of each depicted data point; and
    \textbf{IT}$_3$: the meta-information about the field and data being depicted.
    In Figure \ref{fig:Primitives} (e), monochromic dots convey \textbf{IT}$_1$, while all other plots in the figure convey both \textbf{IT}$_1$ and \textbf{IT}$_2$. Later, when discussing how the primitive designs for different fields are multiplexed together, we will have to consider the need to encode \textbf{IT}$_3$ in order for viewers to de-multiplex visual information belonging to different fields. 
    \item \emph{Data Dimensionality} (DD) --- Although we consider the data of only one field in a primitive design pattern, the field can still be a scalar, vector, or tensor field. We use \textbf{DD}$_1$, \textbf{DD}$_2$, $\ldots$, \textbf{DD}$_k$ to denote 1D, 2D, $\ldots$, $k$-D fields, while \textbf{DD}$_0$ indicates that there is only location information (i.e., \textbf{IT}$_1$). For example, in Figure \ref{fig:Primitives}, we have \textbf{DD}$_0$ in (e), \textbf{DD}$_1$ in (a)$\sim$(d) and (f), and \textbf{DD}$_2$ in (g).
\end{itemize}

Each primitive design in Figure \ref{fig:Primitives} is in fact a representative of a group of primitive design options, since we have not yet decided on the detailed design choices, e.g., which colormap, what shape, and so on. As illustrated at the top of Figure \ref{fig:ThreeSteps}, this is just the first step for exploring the primitive design space coarsely and selecting desired primitive groups.

\subsection{Assigning Primitive Design Options to Data Entities} 
\label{sec:Assign}
After exploring the context-free primitive design patterns, we consider application-specific context and explore the subspace associated with each primitive group. As illustrated in the middle of Figure \ref{fig:ThreeSteps}, for each field, we choose a specific design option from a primitive group by considering application-specific data semantics, user tasks, and perceptive and cognitive merits.

\begin{figure}[hb]
    \centering
    \includegraphics[width=\linewidth]{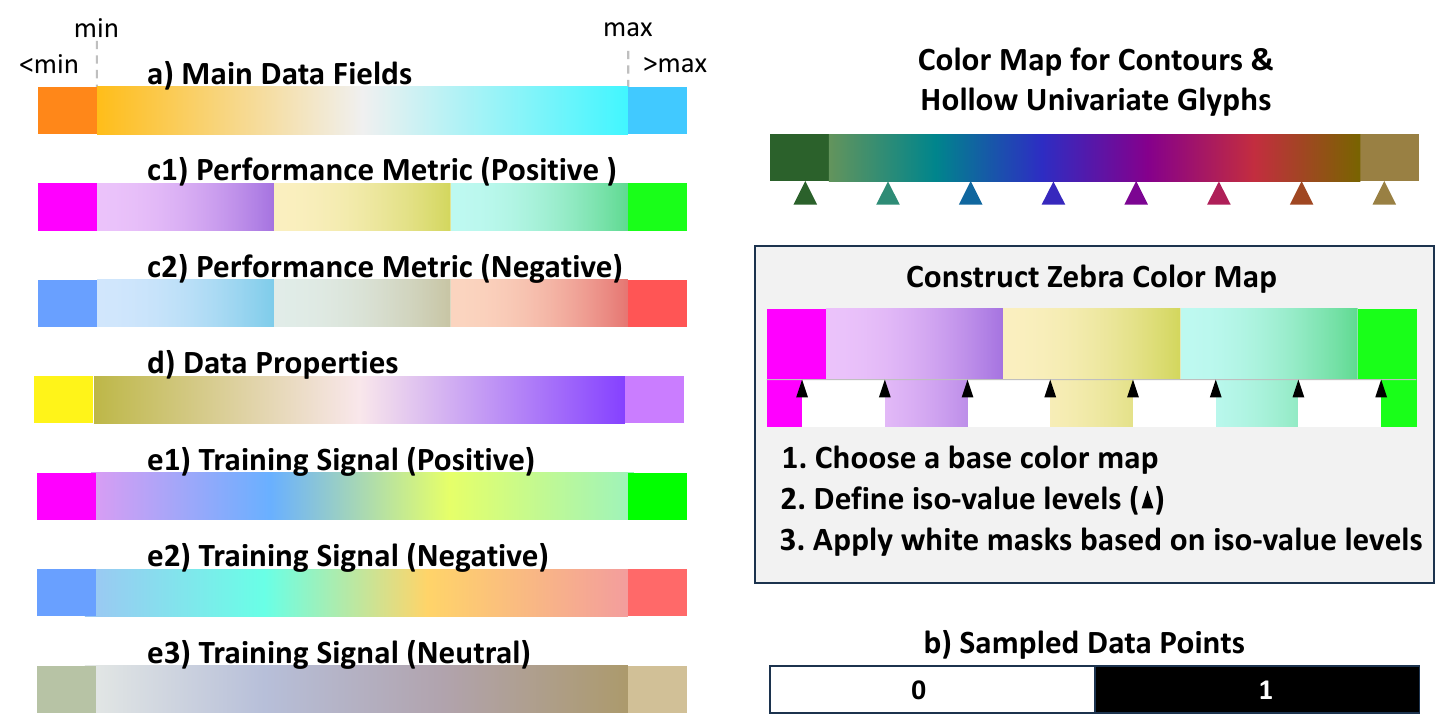}
    \caption{The co-designed default colormaps.}
    \label{fig:colormaps}
\end{figure}

\begin{figure}[h]
    \centering
    \includegraphics[width=\linewidth]{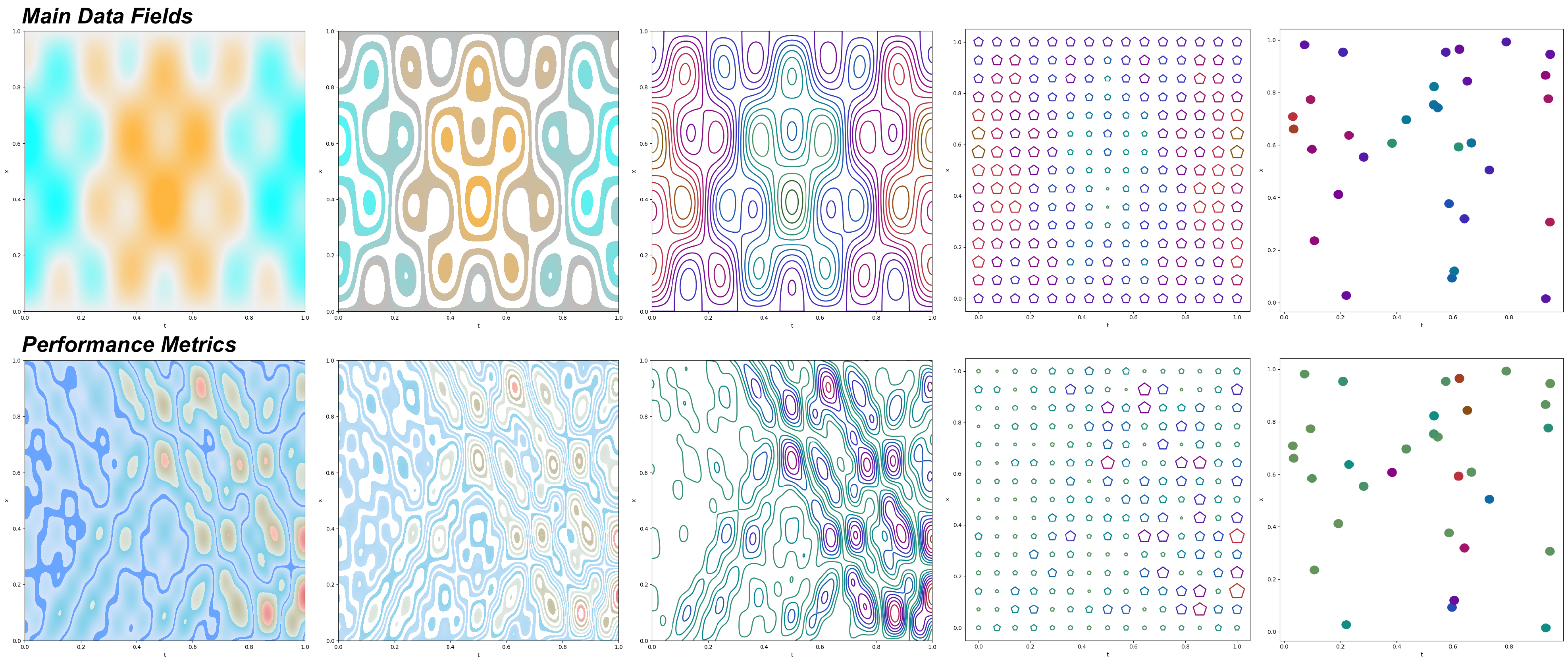}
    \caption{Assigning primitive design options to different categories of fields according to application-specific contexts.}
    \vspace{-5mm}
    \label{fig:Assign}
\end{figure}

As there are over 20 fields in this application (Appendix~\ref{apx:DataSpace}), it is not always beneficial or feasible to assign a unique design option to each field. For example, having 20 different colormaps for 20 fields may incur several cognitive challenges:
(i) becoming difficult to compare fields of a similar nature (e.g., u\_ref vs. u\_pred) using juxtaposition;
(ii) becoming difficult to remember a colormap specification and relying on legends heavily; and
(iii) increasing difficulty when new fields need to be introduced.
We therefore define the following categories of fields for grouping fields of similar data semantics, facilitating juxtapositional comparison for fields in the same category, while avoiding too many variations: 
\begin{enumerate}
    \item[a.] \emph{Main Data Fields} -- fields of numerically-simulated and ML-predicated data values (\#1 u\_ref and \#2 u\_pred in Appendix A.
    %
    \item[b.] \emph{Sampled Data Points} -- a binary field indicating if a data point is in the sampled set (\#14 sample\_set).%
    \item[c.] \emph{Performance Metrics} -- This is further divided into two subcategories: 
    \emph{Positive}: the higher the value, the better (e.g., \#6 phys\_eff); and
    \emph{Negative}: the lower the value, the better (e.g., \#3 phy\_loss and \#4 bc\_loss).
    \item[d.] \emph{Data Properties} -- This includes scalar and vector fields representing additional information about a main data field in (a), e.g., \#19 v\_ref and v\_pred, and \#21 mv\_ref and mv\_pred.
    \item[e.] \emph{Training Signals} -- These metrics provide a view of the model's local learning dynamics at each training sample (e.g., \#7 u\_pred\_grad, \#10 entk, and \#11 phy\_loss\_hessian).
    \item[f.] \emph{ML Hyper-parameters} -- Currently these are not defined in terms of $x$ or $t$, e.g., \#18 lr. However, they can potentially be defined as field functions in the future and be multiplexed with any field in categories (a)$\sim$(e).
\end{enumerate}

The detailed assignment of specific design options requires a good understanding of data semantics (e.g., for categorization) and task requirements (e.g., observational precision and common comparison actions). The different encoding assignments formulated through multiple co-design iterations involving both VIS researchers and domain experts will be detailed in Section \ref{sec:CoDesign}.

Figure~\ref{fig:colormaps} shows the colormaps for categories (a)$\sim$(e), which were co-designed with domain experts (see Section  \ref{sec:CoDesign}).
For category (b), since a set of sampled data points is a binary field (\textbf{DD}$_0$), we use black dots for sampled points, and white for others.
After many experiments in the co-design process, we decided to use a single default colormap with more hue variation for all contours.   
Figure~\ref{fig:Assign} illustrates the colormaps assigned to example scalar fields in categories (a) and (c).

The design of these colormaps also serves several purposes in encoding \emph{meta-information}.
(i) \textit{Field identification.} By designing two-band, three-band, or four-band colormaps for different categories, each colormap becomes a recognizable visual signature. In multiplexed results, observers who recognize a colormap pattern can quickly infer that the category of a data field is being shown without relying heavily on a complex legend; (ii) \textit{Semantic alignment.} Each colormap reflects domain-relevant meaning. For example, for the negative performance metrics and negative training signal groups, smaller values indicate better performance. Assigning cooler hues (blue) to lower values, and warmer hues (red) to larger values, aligns naturally with domain experts' interpretation of error magnitude;
(iii) \textit{Multiplexing separability.} Colormaps must remain distinguishable when layered. 
When primitives are multiplexed, intensity differences help separate layers and reduce perceptual interference. Dense primitives such as heatmaps and zebra maps, which typically occupy the bottom layer, use light colormaps, whereas primitives that commonly appear in upper layers (contours and univariate glyphs) use dark colormaps.

In addition to color, shape can also be used to encode meta-information. Assigning different shapes to different data categories enables rapid identification of the category of a field in multiplexed views, complementing the role of colors.

\subsection{Analyzing Multiplexing Effectiveness} 
\label{sec:Analyze}
After primitive design options are selected, it is necessary to analyze their suitability when primitive designs are multiplexed together. When one needs to create a multiplex plot for $k$ scalar fields, one may choose a primitive design for each field from $g$ groups of primitive visual designs. Since the designs from the same groups can be used for different fields and different ordering of the designs will likely lead to different appearances in multiplexed plots, there are up to $g^k$ possible combinations in the design space of multiplexed plots. In this work, we exhaustively examine all combinations for seven primitive designs in Figure \ref{fig:Primitives} ($g=7$) and two layers ($k=2$). We reason the design space for $k$ layers based on the analysis for $k-1$ layers.

\begin{table}[t]
\newcommand{\bad}{\textcolor{WildStrawberry}{\textbf{X}}}
\newcommand{\good}{\textcolor{ForestGreen}{\textbf{\checkmark}}}
\newcommand{\ok}{\textcolor{Dandelion}{\textbf{!}}}

\centering
\caption{Effectiveness of multiplexing combinations for two fields. Each cell indicates whether a given top-bottom primitive pairing is effective (\good), ineffective (\bad), or conditionally effective (\ok). Letter annotations correspond to the example images shown in Figure~\ref{fig:2dEffective}.}
\label{tab:DStwofield}
\begin{tabular}{@{}c|ccccc@{}}
\hline
\multicolumn{1}{r|}{\textbf{top $\blacktriangleright$}}
    &  
    & \textbf{zebra} 
    &  
    & \textbf{size} 
    & \textbf{color} \\[-1mm]
\textbf{$\blacktriangledown$ bottom}
    & \textbf{heatmap} 
    & \textbf{map} 
    & \textbf{contour} 
    & \textbf{glyph} 
    & \textbf{glyph} \\
\hline

\textbf{heatmap} 
    & \bad      
    & \bad 
    & \good (a)     
    & \good (b)
    & \good (c)     
    \\

\textbf{zebra map} 
    & \bad 
    & \bad     
    & \good (d)
    & \good (e)
    & \good (f)  
    \\

\textbf{contour} 
    & \bad      
    & \bad      
    & \bad      
    & \ok\  (g, j)
    & \good (i)     
    \\

\textbf{size glyph} 
    & \bad 
    & \bad 
    & \ok\  (h)
    & \ok\  (k)
    & \good (l)     
    \\

\textbf{color glyph} 
    & \bad     
    & \bad     
    & \good     
    & \good 
    & \bad       
    \\

\hline
\end{tabular}
\vspace{-0.12in}
\end{table}

\noindent\textbf{Multiplexing Design Space for Two Fields.}
When multiplexing two scalar fields, the choice of primitives and their layering order influences the perceptual clarity of the resulting visualization. Table~\ref{tab:DStwofield} summarizes the effectiveness of pairing different primitives as bottom-top combinations.
\textcolor{black}{The assessment of effectiveness is informed by the information-theoretic cost-benefit analysis, and further supported by validation through a collaborative session with both visualization and domain experts, who examined different combinations and provided perceptual feedback.}

Both the heatmap and its variant, the zebra map, are fundamentally unsuitable as top-layer designs. If transparency were used for the top layer, the colors of both layers would be significantly distorted. If the top layer were fully opaque, there would be too much occlusion. Meanwhile, due to their high information density (\textbf{ID}$_1$), they are highly effective backdrops for medium- and low-density primitives such as contours, size glyphs, and color glyphs.

Contours encode a scalar field through a set of non-intersecting isolines, which offer a more accurate perception of values on isolines than heatmaps while allowing viewers to interpolate the missing information in the empty space. Therefore, contours over a heatmap or zebra map maintain clear separability between the two fields. As shown on the right of Figure \ref{fig:teaser}, the latter combination is particularly effective as the perception of the isolines in both fields is more accurate than heatmaps. The regions (a) and (b) in Figure \ref{fig:teaser} exemplified the merit of this multiplexing design.  

In contrast, when two contour plots are multiplexed, visual interference between the two plots can be serious. For example, the ground truth u\_ref and ML model prediction u\_pred usually have similar contour lines, some of which may intersect or run in parallel nearby. Although reducing the number of isolines may lessen such interference, we considered the combination problematic.

Placing size glyphs over a heatmap or a zebra map is effective in some cases and ineffective in others. When they are placed over contours, the smaller glyphs may easily be missed. In the co-design process, domain experts suggest to provide a UI mechanism that allows users to filter out small values when size glyph is used. In terms of hollow shapes (g) and filled shapes (h), domain experts expressed differing preferences. In terms of the shapes of size glyphs, domain experts offered a consistent view. Shapes with strong angular features, e.g., triangles, squares, and pentagons, were easier to discriminate from smooth contour patterns (j), while circles and near-circular polygons (octagons, decagons) tended to blend visually with isolines, exhibiting poor separability. 

Multiplexing two size-glyph designs is feasible if the glyphs belonging to different layers can be distinguished easily, e.g., via shape difference or color difference. Domain experts noted that such a combination is particularly helpful for studying \textit{correlation patterns}: if both sets of shapes vary in size with similar spatial trends, the fields represented are likely positively correlated, while divergent patterns indicate negative correlation.

In general, color glyphs can be superimposed effectively on all other primitives, making them one of the most commonly used multiplexing options. However, multiplexing two different color glyph layers is generally problematic due to poor separability. Nevertheless, they can be used in multivariate glyphs discussed below.

\begin{figure*}[t]
    \centering
    \includegraphics[width=0.82\linewidth]{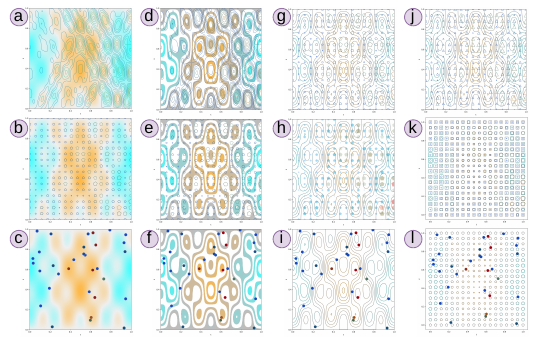}
    \caption{Examples of two fields multiplexed results. In all images, \textit{u\_ref} is the bottom field, with either \textit{residuals} (first two columns) or training-sample \textit{data\_loss} (last column) as the top field.}
    \vspace{-4mm}
    \label{fig:2dEffective}
\end{figure*}

\begin{figure}[t]
    \centering
    \includegraphics[width=\linewidth]{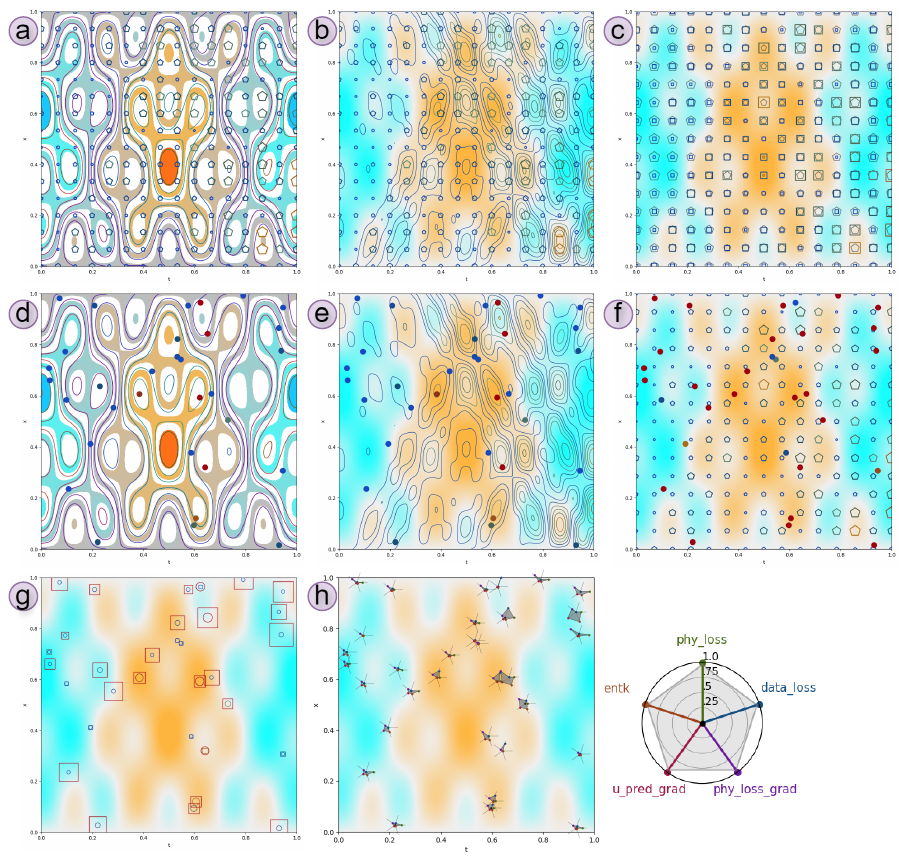}
    \caption{Examples of three or more fields multiplexed results.}
    \vspace{-4mm}
    \label{fig:3dEffective}
\end{figure}

\noindent\textbf{Multiplexing Design Space for Three or More Fields.}
Instead of examining all 125 possible combinations of primitive designs, we can analytically exclude any combination containing ineffective combinations in Table \ref{tab:DStwofield}. Among 125 possible combinations, 27 remain. Among these, 15 combinations are conditionally effective as they involve size-size and size-contour combinations. After careful checking, the remaining 12 were found to be effective.  

Although multiplexing layers of univariate color glyphs is ineffective, we experimented with a type of multivariate glyphs, where color circles for different layers at the same position are combined into a radial graph. This, in fact, is Type H multiplexing ``shift a visual channel'' or Type J ``visual language'' in \cite{Chen:2014:CGF}. In the co-design process, domain experts were particularly impressed by such radial glyphs, as they can be used to display different types of training signals over a heatmap, zebra map, or contour plot.

Importantly, radial glyphs can be used for multiplexing more than three scalar fields. Figure \ref{fig:3dEffective} shows some examples of multiplexing three or more layers.

\section{\textcolor{black}{Co-Design as Method Instantiation and Validation}}
\label{sec:CoDesign}

To construct and explore a context-specific multiplexing design space, it is necessary to have sufficient knowledge about the context. For our context, the knowledge of ML workflows in general, and ML models \textcolor{black}{and derived data} for PDEs in particular, is important.
\textcolor{black}{To instantiate and examine the proposed design-space construction method, we engaged domain experts throughout the process. Importantly, this collaboration is not framed as a conventional design study in which requirements are derived primarily by domain experts~\cite{Sedlmair:2012:TVCG}. In our example, this was not effective as domain experts could not articulate their requirements easily due to insufficient knowledge about visualization.}

\textcolor{black}{Instead, following initial effort for gathering user requirements, we used the theory-driven method as described in Section \ref{sec:Theory} to identify problems in their current workflow, and reason about potential solutions.
We then engaged with domain experts in a co-design process, where domain expertise helped contextualize and refine the design space, while domain experts became more engaged in expressing their preference in terms of visualization.
In this work, the role of domain experts arises from three aspects. First, domain context constrains the otherwise vast multiplexing design space into a manageable, application-relevant subspace. Second, perceptual and preference-sensitive design decisions, such as colormap selection and visual separability, require human judgment that cannot be fully derived from theoretical analysis alone. Third, practical considerations, such as how users may explore alternative designs, inform the design of an interface prototype that supports navigation within the constrained design space.
In contrast to traditional design studies, where requirement analysis is largely driven by expert interviews and iterative prototyping, our requirements analysis largely relies on the cost-benefit ratio to identify symptoms, analyze causes, formulate remedies at the high-level, and consider potential side-effects. Our design process relies on the three-step method described in Section \ref{sec:Method} to discover 
candidate design options systematically, as well as the co-design process to refine design decisions.
In the following, we describe this collaboration, which instantiates and validates our approach.}

\begin{figure}[t]
    \centering
    \includegraphics[width=\linewidth]{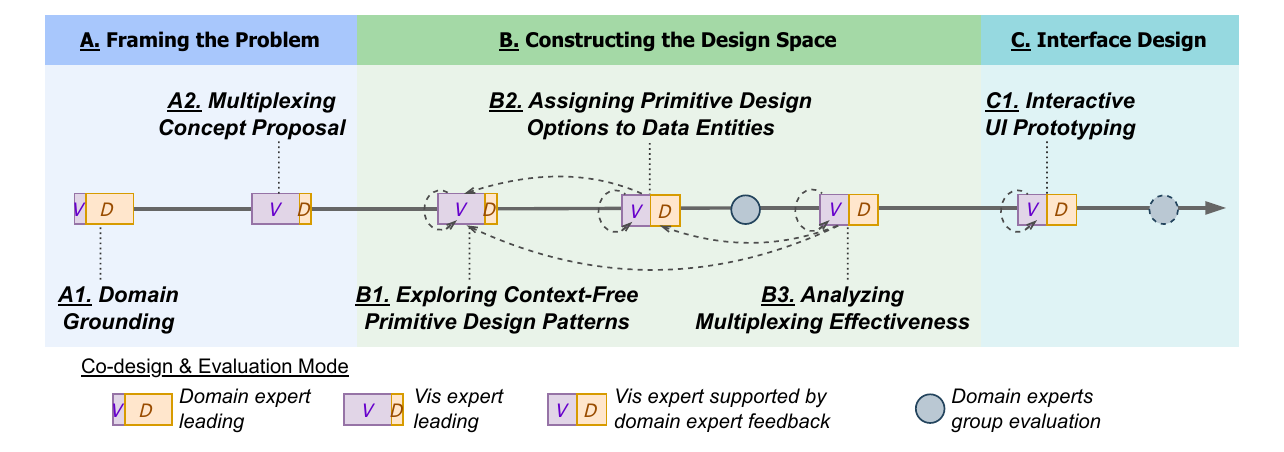}
    \caption{The timeline of co-design and evaluation stages.}
    \vspace{-4mm}
    \label{fig:evaluation}
\end{figure}

We adopted a longitudinal approach, involving domain experts weekly in every phase of the work, including the three steps outlined in Section~\ref{sec:Method}. 
Figure~\ref{fig:evaluation} illustrates the major milestones of this process. Our frequent continuous engagement includes framing the problems in ML workflows as well as in visual designs, evaluating design options in exploring different aspects of a design space, querying about data semantics and user tasks, discussing trade-offs among conflicting design decisions, and so on. In some cases (e.g., identifying primitive designs), the processes were led by VIS researchers, and in other cases (e.g., evaluating the effectiveness of different multiplexed plots), the processes focused on domain experts. Such changes of focus shifted fluidly depending on the topics under discussion. 

Throughout this process, a PDE modelling expert (2nd author) and an ML expert familiar with visualization concepts (3rd author) worked closely with the VIS researchers. The 3rd author also acted as a liaison, facilitating communication across domains and visualization perspectives~\cite{Simon:2015:EuroVis}. Together, the experts contributed domain knowledge, PDE data, contextual interpretation, and iterative feedback that informed and refined the visualization designs. The detailed co-design process is described in Section~\ref{sec:codesign}.

We also complemented the co-design processes with a semi-structured group discussion involving six additional domain experts. This session served as a validation checkpoint: confirming the generality of earlier observations, mitigating the risk of over-reliance on a small collaborator set, and providing broader feedback to inform subsequent refinements of the multiplexing strategies. The procedure and findings are presented in Section~\ref{sec:evalution-discussion}.

\subsection{Longitudinal Co-Design}
\label{sec:codesign}

The co-design and feedback process took the form of weekly meetings supplemented by continuous email-based discussions. This longitudinal collaboration unfolded across three major phases and can be further understood through the six milestone stages illustrated in Figure~\ref{fig:evaluation}. The insights derived from the three key stages in Phase B are reported in Section~\ref{sec:Method}. 

\textbf{Co-Framing the Problem.} 
Initial discussions focused on understanding the modeling workflow, the structure of PDE data, and the performance metrics used during training. Domain experts contributed significantly during the domain-grounding stage (Figure~\ref{fig:evaluation} \textit{A1}) by introducing their workflow. Through iterative questioning, visual probing, and reflective discussion with visualization experts, they gradually articulated two key challenges: 
1) Lack of visual support to inspect local behavior across epochs.
Despite routinely producing field data and multiple loss terms (e.g., physics loss, boundary-condition loss, data loss), practitioners typically rely on global summaries such as learning curves. Existing tools provide no mechanism for examining how these metrics vary locally over the $(x,t)$ region during training. 
2) Difficulty comparing multiple fields defined on the same region.
Practitioners need to jointly examine reference fields, predictions, and loss metrics in the same region. Common approach, such as side-by-side juxtaposing heatmaps of different fields, fails to support fine-grained comparison, as spatial alignment and local correspondence are visually fragmented.

These challenges established the need for a visual approach capable of overlaying heterogeneous fields on the same domain while maintaining interpretability. During this stage (Figure~\ref{fig:evaluation} \textit{A2}), visualization experts took the lead by proposing multiplexing as a potential solution. The domain expert distilled a list of commonly used data fields and metrics, which informed the subsequent construction of the multiplexing design space.

\textbf{Co-Constructing the Design Space.} 
Given the large number of data fields and metrics provided by the domain expert, a bespoke design for each field was neither scalable nor conceptually coherent. We therefore collaboratively conducted a data abstraction that categorized the domain data into six major types. During this stage (Figure~\ref{fig:evaluation} \textit{B1}), the visualization experts led the development of six primitive design patterns, excluding the (c) \textit{zebra map} in Figure~\ref{fig:Primitives}. 
These primitive designs served as basic pillars of the design space. 
We then created, evaluated, and refined multiplexed visualizations across combinations of these primitives with the two domain collaborators, and supported by the group discussion (detailed in Section~\ref{sec:evalution-discussion}). 
This iterative co-design process in \textit{B2} and \textit{B3} surfaced considerations of 
\textit{How can multiple fields be layered without being visually overwhelmed?
And how should color mappings be designed so that they encode data semantics while remaining distinguishable when overlapped?}

A substantial portion of the co-design effort focused on iterating over color-mapping strategies. Over the course of the collaboration, we iteratively tested four colormap sets, comprising over 60 individual colormaps. The resulting final designs are presented in Figure~\ref{fig:colormaps}. 
Our early approach aimed to preserve color-semantic consistency by assigning each data category a dedicated colormap family. As a result, each design set contained seven groups of colormaps: one for \textit{main data fields}, two for \textit{performance metrics} (positive and negative), one for \textit{data properties}, and three for \textit{training signals} (positive, negative, and neutral). To further support layer differentiation in multiplexed views, we created both light and dark variants for every group, enabling combinations such as using a darker version for contours and glyphs, and a lighter version for heatmaps.

However, during co-design evaluations, we observed that intensity-based separation alone was not sufficient. When the same data type appeared in multiple layers, for example, a heatmap showing u\_ref overlaid with contours of u\_pred, the \textit{darker–lighter} pairing became visually ambiguous, and the color of the upper contour layer was easily lost. We realized that the just-notifiable difference is affected by both intensity and thickness of the contour lines \cite{AlbersSzafir:2018:TVCG}.
This limitation motivated a shift toward using a separate colormap (the top-right colormap in Figure~\ref{fig:colormaps}), featuring a richer hue range and more variation of luminance, for contours and hollow shape glyphs when placed over a heatmap, thereby improving their perceptual separation from the underlying layer.

Our initial colormap designs used fully continuous gradients, which aligned with the numerical and continuous nature of these data fields. However, the domain experts emphasized that for metrics such as data\_loss, analytical attention is often placed on broader value regions rather than precise numeric readings. For example, large loss values signal failure cases and should stand out clearly (e.g., a red band), and small values are comparatively less critical (e.g., blue).
To support this domain-driven interpretation, we designed discontinuous color bands within an otherwise continuous colormap (e.g., Figure~\ref{fig:colormaps} c1\&c2). These perceptual breaks allow users to rapidly determine which band a value belongs to when viewing multiplexed layers, while the gradient within each band still preserves fine-grained detail. This combination of categorical segmentation and continuous interpolation is especially effective for observing coarse banding and fine visual patterns. 

As the co-design progressed, one recurring analytical need became especially salient: comparing the reference field with the predicted field. This is always encoded using the Heatmap primitive for u\_ref, and Contour for u\_pred. However, both the visualization and domain experts found that, although this layering conveyed the general trend, it remained difficult to precisely locate where the prediction aligned well with the reference and where it deviated.
To improve the clarity of this comparison, visualization experts experimented with enabling the heatmap to convey iso-value structure that could be visually aligned with the contour lines. Building on the original colormap for the heatmap, we inserted white masks at intermediate value intervals corresponding to the contour line iso-values. These white interruptions created an alternating light-dark banding pattern -- what we termed the zebra colormap (Figure~\ref{fig:colormaps} centre-right section). When applied to the heatmap, this pattern produced the hybrid visualization -- the zebra map.

Both domain and visualization experts agreed that the zebra map allowed them to detect where u\_pred diverged from u\_ref along matching iso-value boundaries, making regions of large discrepancy stand out (the case \textcircled{b} in Figure~\ref{fig:teaser}). 
The domain collaborator who provided the data offered domain-specific insights: \textit{``predicted and reference contours mismatched in the bottom right corner due to high residual error. Upon further investigation of the per-sample physics loss, I see it is high in those sampling regions. It indicates that the physics-informed neural network fails to enforce the governing equation locally. I hypothesize that enforcing more weight on the physics loss component or using a different adaptive weighting method could reduce the error.''}
The zebra map proved useful for fine-grained comparison between similar fields, and was therefore added as an additional primitive design pattern (Figure~\ref{fig:Primitives} c).

Domain experts raised an important question during the co-design processes, regarding the comparability of colormaps when examining model performance across different epochs. For instance, the residual range at epoch 20,000 may span $[-0.2,0.2]$, whereas at epoch 40,000 it may narrow to $[-0.1,0.1]$. If the colormap automatically adapts to the data range of each field (a common default in many plotting libraries), the same color no longer corresponds to the same numerical value across epochs, making cross-epoch comparison unreliable.
We concluded that the main colormap should have a fixed numerical range, rather than adjusting its $v_{\min}$ and $v_{\max}$ per dataset. The fixed range can be determined by the user based on domain knowledge (e.g., defining what constitutes a ``normal'' residual magnitude). Values falling outside this prescribed range are mapped using dedicated under/over colors (the extreme-value color blocks at the two ends of the colorbars in Figure~\ref{fig:colormaps}), visually marking exceptions beyond the expected bounds. This strategy ensures (1) comparability throughout a workflow via a stable reference -- color mapping, and (2) full coverage from $-\infty$ to $+\infty$ via under/over colors, and (3) clear visual highlighting of extreme values outside the user-defined meaningful range.

When seeking domain experts' feedback on the effectiveness of different multiplexed plots, the expert who provided the PDE data found the radar glyph (Figure~\ref{fig:3dEffective} h) useful: \textit{``it clearly shows various loss terms. I see the loss-balance behaviour of that particular iteration, and it shows the relative contribution of each loss term.''} Building on this observation, she expressed interest in future ML work on exploring whether adjusting those loss term weights can improve training results. 

Through multiple cycles of co-design and refinement, we produced a series of multiplexed visualizations using real PDE data, experimenting with different colormaps and primitive variants, and assessed their effectiveness together with the domain experts. The effective results are summarized in Section~\ref{sec:Analyze}.

\textbf{Interface Co-Design.}
During the design space construction phase, we observed that different primitive combinations and variants carried distinct merits and demerits. Preferences may vary across observers, which highlighted the need for interactive controls that allow users to adjust (1) primitive design assignments, (2) colormaps, including threshold for extreme values, (3) layering orders, and (4) fine-grained encodings, such as the shape, hollow or filled, line style, iso-value levels, and so on, according to analytical goals and subjective preferences. 
In addition, the use of multivariate glyphs requires an accompanying glyph legend placed adjacent to the multiplexed result to support interpretation. 
\textcolor{black}{These considerations motivated a co-design phase for a conceptual user interface prototype (Figure 9 Phase C).}
Earlier discussions between domain experts and VIS researchers led to a converged set of default multiplexing strategies, meaning that when multiple fields are selected, an effective combination of primitives, variants, and colormaps is applied automatically. The UI should allow users to (1) select data from a specific training iteration, (2) adjust the multiplexing defaults, and (3) interactively explore the multiplexed results.

\subsection{Semi-Structured Group Discussions}
\label{sec:evalution-discussion}
After implementing a collection of multiplexed visualizations, we conducted a 90-minute semi-structured group discussion. Using convenience sampling through our domain collaborators, we recruited 6 experts: 3 in PDE or physics-ML, 1 in robotics ML, and 2 in general ML. The two collaborators also attended as facilitators. \emph{This discussion session followed the University of Oxford's guidelines on empirical studies and expert consultation. It did not require a formal approval as a consultation session.}  

\noindent \textbf{Approach.}
The session, conducted via Microsoft Teams, followed a three-part structure. 
(1) Experts were invited to introduce their background and analytical practices. This phase served both as an icebreaker and as a way to assess whether the challenges surfaced during co-design were broadly representative. 
(2) We presented a suite of multiplexed visualizations generated from real PDE data, demonstrating how multiple fields can be displayed simultaneously. Experts were asked to comment on their interpretability and usefulness. 
(3) Experts reflected more broadly on multiplexing: how well the approach aligns with their analytical needs, and what improvements could enhance its practical value. This open-ended discussion elicited preferences and concerns related to color mapping, layer separability, annotation requirements, and other factors shaping the usability of multiplexed results.

\noindent \textbf{Experts' Feedback.}
Experts confirmed that existing tools in their workflows mainly expose global performance indicators (e.g., learning curves) and provide limited support for examining local variations in loss terms or intermediate fields. Some noted that they seldom inspect local metrics, not due to lack of relevance, but because current tooling makes such analysis cumbersome. These remarks echoed our co-design findings and suggested that the need for local inspection extends beyond the PDE context.

Experts responded positively to the multiplexing approach. They agreed that the multiplexed results provided clearer comparative insight than juxtaposition when fields need to be examined over the same domain: \textit{``I tend to go for the overlaid version, it's more understandable and more aesthetic.''}
Experts also expressed interest in applying multiplexing to their own works. The robotics practitioner commented that multiplexing would allow him to \textit{``compare different depth information on a spatial map rendered as an RGB image,''} suggesting utility beyond the PDE context.

Experts also made constructive suggestions for improving some multiplexed designs shown. Many of the concerns were about color mapping. In some examples, we used a single-hue sequential colormap for residuals. Experts, however, expressed a preference for a rainbow-style colormap and found the mapping of low residual values to light colors unsatisfying, noting that these values became \textit{``hard to read and easy to miss.''} 
Experts also commented on the limitations of multiplexing when too many dense layers are overlaid, observing that interpretability decreases once more than three visually heavy layers are combined. They also suggested incorporating supportive visual cues, beyond conventional legends, to make it immediately clear which fields were being shown. These comments informed the subsequent co-design cycles.

\section{Discussions and Conclusions}
\label{sec:Conclusions}
Multiplexing is a natural asset of visualization. In this paper, we proposed a method for exploring the design space of multiplexing in the context of a specific application. Since the overall design space of multiplexing is too vast to be explored systematically, adopting a divide-and-conquer approach offeres a feasible step toward advancing visualization design methods, while developing our ability to use multiplexing as an advantageous asset in practical visualization solutions.

We recognize that the subspace for multiplexing 2D scalar fields occupies only a small portion of the entire design space of multiplexing. However, we believed that our method, together with the co-design and evaluation activities, could be used to explore other context-specific design spaces of multiplexing. Examples of such subspaces include multiplexing data associated with multiple point clouds, multiple maps, multiple 3D surfaces, multiple graphs, and so on. As more of these subspaces are explored by VIS researchers, we will likely gain substantial new insights and a deeper understanding of multiplexing and its design space. Hopefully, there will be proposed methods for a large portion of the multiplexing design space, or better, the entire space.

Our co-design process revealed a general tendency: the more layers one attempts to multiplex, the lower the average information density across those layers. From heatmap, to zebra map, contour, and then univariate glyphs, the direction of travel is essentially an increase of alphabet compression (i.e., entropy reduction or information loss). When multiplexing more than three layers, one cannot help considering multivariate glyphs as part of the multiplexing strategy. Prior work has shown that glyphs can encode as many as 20 variables \cite{Duffy:2015:TVCG}, making the multiplexing design space connected to the glyph design space \cite{Borgo:2013:CGF,Hsieh:2025:TVCG}. We hope that future research will study this connection in depth.

The author team of this work will continue to provide visualization support to ML workflows. In particular, as multiplexed visualizations allow domain experts to observe more local statistics, the team plans to identify local statistics that can inform the internal computational processes for improving ML models automatically.    


\acknowledgments{%
  This work is part of the VIS4ML4HD project funded by UK Research and Innovation (EPSRC: EP/X029557/1).%
}

\bibliographystyle{abbrv-doi-hyperref-narrow}
\bibliography{references}

\newpage
\appendix 
\crefalias{section}{appendix}

\begin{center}
\large
APPENDIX\\[1mm]
\LARGE\noindent
\textbf{\textsf{A Multiplexing Design Space:\\
Theory, Method, and Application}}\\[2mm]
\normalsize
Y. Xing$^1$, A. Farea$^2$, S. Khan$^3$, \& M. Chen$^1$\\[1mm]
$^1$University of Oxford, UK\\
$^2$Istanbul Technical University, Turkiye\\
$^3$Science and Technology Facilities Council (STFC), UK
\normalsize
\end{center}

\section{\textbf{Data Space of the Case Study}}
\label{apx:DataSpace}
As described in Section \ref{sec:Context}, this work focuses on an application case study, where ML models are being developed to approximate numerical solvers of partial differential equations (PDEs). In particular, we consider 1D PDEs, and their numerical solutions are typically represented as 2D scalar fields in the form of
\[
u(x, t),
\quad x \in \mathbf{X} = \{ x_1, x_2, \ldots, x_{nx} \};
\quad t \in \mathbf{T} = \{t_1, t_2, \ldots, t_{nt} \}
\]

During an ML workflow for training and testing such a model, many types of derived variables can be computed. These derived variables are described below. Most of them are related to the domain of $\mathbf{X} \times \mathbf{T}$, hence they can be represented as field data. The exceptions are \#15, \#16, \#17, and \#18, which are normally not set in relation to $x, t$. However, in computational fluid dynamics, numerical models can have spatially- or temporally-localized settings for parameters and boundary conditions. There is no reason why this would not happen in future ML workflows. Hence, we consider \#15, \#16, \#17, and \#18 are potential field data.  

\noindent
\#1. \emph{Reference numerical solution of a PDE}
\begin{itemize}
    \item \textbf{Abbreviated Label:} u\_ref.
    \item \textbf{Function:} $u(x, t), x \in \mathbf{X}, t \in \mathbf{T}$.
    \item \textbf{Comment:} This is a main data field in an ML workflow, and is usually used as the ground truth for interim sample-based testing during training and for evaluating a full predicted solution during testing.  
\end{itemize}

\#2. \emph{Solution predicated by an ML model}
\begin{itemize}
    \item \textbf{Abbreviated Label:} u\_pred.
    \item \textbf{Function:} $u_\theta(x, t), x \in \mathbf{X}, t \in \mathbf{T}$.
    \item \textbf{Comment:} This is a main data field in an ML workflow, and it typically represents a full predicted solution during testing.
\end{itemize}

\#3. \emph{PDE residual at (x, t)}
\begin{itemize}
    \item \textbf{Abbreviated Label:} phys\_loss.
    \item \textbf{Function:} $\mathcal{L}_{phy} = \mathcal{L}(\mathcal{D}[u_\theta(x, t); \boldsymbol{\alpha}] - f(x, t))$.
    \item \textbf{Comment:} ) It shows how well the model satisfies the governing PDEs.  
\end{itemize}

\#4. \emph{Data loss at (x, t)}
\begin{itemize}
    \item \textbf{Abbreviated Label:} data\_loss.
    \item \textbf{Function:} $\mathcal{L}_{data} = \mathcal{L}(u(x, t) - u_\theta(x, t))$.
    \item \textbf{Comment:} The difference between the reference and the predicted solution. This loss is not used for optimization purposes, especially in forward problems. 
\end{itemize}

\#5. \emph{Boundary/Initial condition loss}
\begin{itemize}
    \item \textbf{Abbreviated Label:} bc\_loss.
    \item \textbf{Function:} $\mathcal{L}_{bc} = \mathcal{L}(u_{bc}(x, t) - u_{\theta_{bc}}(x, t))$.
    \item \textbf{Comment:}  It shows the satisfaction of other loss terms, such as initial and boundary conditions.  
\end{itemize}

\#6. \emph{Physics-data conflict score}
\begin{itemize}
    \item \textbf{Abbreviated Label:} phys\_eff.
    \item \textbf{Function:} $\Psi(\text{bc\_loss},\text{phys\_loss})$.
    \item \textbf{Comment:} A score that measures conflict between physics enforcement and other loss terms.
\end{itemize}

\#7. \emph{Gradient of the model output with respect to its trainable parameters.}
\begin{itemize}
    \item \textbf{Abbreviated Label:} u\_pred\_grad.
    \item \textbf{Function:} $\nabla_\theta u_{\theta}(x,t)$.
    \item \textbf{Comment:} It can show the neural network’s output sensitivity to parameter changes.  
\end{itemize}

\#8. \emph{Gradient of the physics loss with respect to the neural network parameters.}
\begin{itemize}
    \item \textbf{Abbreviated Label:} phy\_loss\_grad.
    \item \textbf{Function:} $\nabla_\theta \mathcal{L}_{phy}( u_{\theta}(x))$.
    \item \textbf{Comment:} It diagnoses sensitivity to parameter changes, conflict with other losses, dominance, or underweighting of the physics term.  
\end{itemize}

\#9. \emph{Gradient of the boundary/initial condition loss with respect to the neural network parameters.}
\begin{itemize}
    \item \textbf{Abbreviated Label:} bc\_loss\_grad.
    \item \textbf{Function:} $\nabla_\theta \mathcal{L}_{bc}( u_{\theta_i}(x_i))$.
    \item \textbf{Comment:} It helps diagnose learning dynamics and imbalance in the boundary/initial conditions loss terms.  
\end{itemize}

\#10. \emph{Empirical neural tangent kernel}
\begin{itemize}
    \item \textbf{Abbreviated Label:} entk.
    \item \textbf{Function:} $\Theta_{ij} = \nabla_\theta u_{\theta_i}(x_i,t_k; \theta)^T \nabla_\theta u_{\theta_j}(x_j,t_k; \theta)$.
    \item \textbf{Comment:} A matrix that encodes how outputs co-vary under parameter changes.  
\end{itemize}

\#11. \emph{Hessian of the physics residual loss with respect to the neural network parameters}
\begin{itemize}
    \item \textbf{Abbreviated Label:} phy\_loss\_hess.
    \item \textbf{Function:} $H^{(\mathcal{L}_{phy})}_{ij} = \frac{\partial^2 \mathcal{L}_{phy}(x,t)}{\partial \theta_i \partial \theta_j}$.
    \item \textbf{Comment:} It measures the curvature of the residual loss to parameter changes.  
\end{itemize}

\#12. \emph{Hessian of the boundary/initial condition loss with respect to the neural network parameters.}
\begin{itemize}
    \item \textbf{Abbreviated Label:} bc\_loss\_hess.
    \item \textbf{Function:} $H^{(\mathcal{L}_{bc})}_{ij} = \frac{\partial^2 \mathcal{L}_{bc}(x,t)}{\partial \theta_i \partial \theta_j}$.
    \item \textbf{Comment:} It measures the curvature of the boundary/initial condition loss to parameter changes.
\end{itemize}

\#13. \emph{Density function of the gradient.}
\begin{itemize}
    \item \textbf{Abbreviated Label:} grad\_pdf.
    \item \textbf{Function:} $p(\nabla_\theta \mathcal{L}(x,t))$.
    \item \textbf{Comment:} This can be derived from \#8 or \#9, or summation of both.  
\end{itemize}

\#14. \emph{Sampling points}
\begin{itemize}
    \item \textbf{Abbreviated Label:} sample\_set.
    \item \textbf{Function:} $s(x, t) =$ 1 (if selected) or 0 (otherwise).
    \item \textbf{Comment:} In each epoch, a subset of points is randomly selected from the whole point set $\mathbf{X} \times \mathbf{T}$, upon which $u(x, t)$ and $u_\theta(x, t)$ are defined. In programming, a sample\_set is often defined as a set $S = \{ p_1, p_2, \ldots, p_n \}$, where $p_i$ is a sampled instance of $(x, t), x \in \mathbf{X}, t \in \mathbf{T}$. An alternative definition is a binary scalar field as given in the above bullet point. Hence, the plot in Figure \ref{fig:MLworkflow}(c) is a binary heatmap as well as a 2D scatter plot.
\end{itemize}

\#15. \emph{Types of training data sampling distribution}
\begin{itemize}
    \item \textbf{Abbreviated Label:} data\_dist.
    \item \textbf{Mathematical Symbol:} $(T_\text{dsd})$
    \item \textbf{Comment:} E.g., Sobol, Random, LHS, etc.  
\end{itemize}

\#16. \emph{Neural network trainable parameters}
\begin{itemize}
    \item \textbf{Abbreviated Label:} nn\_params.
    \item \textbf{Mathematical Symbol:} $(\theta)$.
\end{itemize}

\#17. \emph{Neural network parameter initialization scheme}
\begin{itemize}
    \item \textbf{Abbreviated Label:} param\_dist.
    \item \textbf{Mathematical Symbol:} $\text{Init}(\theta)$
    \item \textbf{Comment:} E.g., He, Xavier, etc.  
\end{itemize}

\#18. \emph{Learning rate hyperparameter}
\begin{itemize}
    \item \textbf{Abbreviated Label:} lr.
    \item \textbf{Mathematical Symbol:} $\eta$.
    \item \textbf{Comment:} Learning rate at the current epoch.  
\end{itemize}

\#19. \emph{Estimated normal vectors}
\begin{itemize}
    \item \textbf{Abbreviated Label:} v\_ref, v\_pred.
    \item \textbf{Function:} $\vec{v}(x, t), \vec{v}_\theta(x, t), x \in \mathbf{X}, t \in \mathbf{T}$.
    \item \textbf{Comment:} When $x$ is a 2D point, one considers that an iso-line or iso-curve passes through $x$, and the normal of the iso-line or iso-curve at $x$ can be estimated using the data values of $x$ and its neighbor. Similarly, in 3D, one considers an iso-surface and in a higher dimensional space, one considers an iso-hypersurface. The simplest and most common way to estimate such a normal vector is the central difference method. There are also more complicated methods, e.g., polynomial surface fitting; cross-product of partial derivatives; Sobel, Scharr, and other image gradient filters; and so on. In 1D, the central difference defines a vector at $x_i, t_j$ as:
    \begin{align*}
        dx_i &= \kappa \bigl( u(x_{i+1}, t_j) - u(x_{i-1}, t_j) \bigr)\\
        dt_j &= t_{j+1} - t_j
    \end{align*}
    where $\kappa$ is a scaling factor to moderate the scale difference between $dx$ and $dt$. 
\end{itemize}

\#20. \emph{Estimated traversal directions}
\begin{itemize}
    \item \textbf{Abbreviated Label:} d\_ref, d\_pred.
    \item \textbf{Function:} $\vec{d}(x, t), \vec{d}_\theta(x, t), x \in \mathbf{X}, t \in \mathbf{T}$.
    \item \textbf{Comment:} When a data point $x_i$ represents a physical entity, one may estimate how $x_i$ moves from its current position $(x_i, t_j)$ to a new position $(\mathring{x_i}, t_{j+1})$. Many algorithms for motion estimation can be used, including variants of the optical flow algorithm. In recent years, some ML models have been shown to be effective in motion estimation.  
\end{itemize}

\#21. \emph{Magnitudes of vectors}
\begin{itemize}
    \item \textbf{Abbreviated Label:} mv\_ref, mv\_pred, md\_ref, md\_pred.
    \item \textbf{Function:} $mv(x, t), mv_\theta(x, t), md(x, t), md_\theta(x, t), x \in \mathbf{X}, t \in \mathbf{T}$.
    \item \textbf{Comment:} In many applications, one uses heatmaps to observe the vector magnitudes of a vector field. A field of magnitudes is a scalar field, which can easily be computed from a vector field, such as $\vec{v}(x, t)$, $\vec{v}_\theta(x, t)$,  $\vec{d}(x, t)$, and $\vec{d}_\theta(x, t)$. 
\end{itemize}

\end{document}